\documentclass[12pt,final]{amsart}

\usepackage{graphicx} 
\usepackage[margin=0.75in]{geometry}
\usepackage{amsmath}

\usepackage{verbatim}
\usepackage{amssymb}
\usepackage{amscd}

\newtheorem{lemma}{Lemma}[section]
\newtheorem{proposition}[lemma]{Proposition}
\newtheorem{theorem}[lemma]{Theorem}

\theoremstyle{definition}
\newtheorem{defn}[lemma]{Definition}
\theoremstyle{remark}
\newtheorem{remark}[lemma]{Remark}
\newtheorem{example}[lemma]{Example}

\newcommand{\Z}{\ensuremath{{\mathbb Z}}}
\newcommand{\N}{\ensuremath{{\mathbb N}}}

\newcommand{\R}{\ensuremath{\mathbb R}}

\newcommand{\tr}{\ensuremath{\mathrm {tr}}}

\author{Tatyana Barron}
\address{Department of Mathematics, University of Western Ontario, London Ontario N6A 5B7, Canada}
\email{tatyana.barron@uwo.ca}

\author{Spencer Kelly}
\address{Department of Mathematics, University of Western Ontario, London Ontario N6A 5B7, Canada}
\email{skell86@uwo.ca}

\author{Colin Poulton}
\address{Department of Mathematics, University of Western Ontario, London Ontario N6A 5B7, Canada}
\email{cpoulto@uwo.ca}

\thanks{T.B. acknowledges partial support from the University of Western Ontario Bridge grant Proposal ID 56499. S.K. and C.P. were partially supported by the University of Western Ontario USRI program}

\title{Signals as submanifolds, and configurations of points}
\begin{document}
\sloppy

\maketitle

\noindent {\bf Abstract.} 
For the purposes of abstract theory of signal propagation, a signal is a submanifold of a Riemannian manifold. 
We obtain energy inequalities, or upper bounds, lower bounds on energy  in a number of specific cases, including parameter spaces of Gaussians and spaces of configurations of points. 
We discuss the role of time as well as graph embeddings.  

\
  
\noindent {\bf MSC 2020}: 53Z05; 94A12

\

\noindent {\bf Keywords}: Riemannian manifold, energy, information, Gaussian

\section{Introduction} 

This paper follows \cite{barron:gs1}, where a signal, in an abstract sense, was defined as a specific type of a cobordism in a Riemannian manifold. Here we use a definition which essentially allows us to consider an arbitrary submanifold of a Riemannian manifold, with very mild restrictions. In section \ref{sec:signals}, we discuss various definitions of the energy of a signal and obtain upper bounds on energy for particular types of signals: 
Theorem \ref{th:enbound}. Expanding on the remarks in \cite{barron:gs1}, we discuss time, noise, and Fourier transform. We explain that it makes sense to talk about "$m$-dimensional time" with an integer $m\ge 2$. 
We consider specific Riemannian manifolds 
that appear in applications.     
 In section \ref{sec:gau}, we obtain  a positive lower bound on the energy of a $1$-dimensional signal in the space of Gaussian distributions 
 on $\R^n$ (equipped with the product metric, not with the Fisher information metric): Theorem \ref{th:e2}. Let $n\ge 2$ be an integer. 
 In sections \ref{sec:completeg} and \ref{sec:graphemb}, we consider spaces $C_n(M)$ of configurations of $n$ points in a connected Riemannian manifold $M$
 \cite{atiyah}. When $M$ is a $3$-dimensional spherical shell in $\R^3$ (a region between two $2$-dimensional spheres centered at the origin), we obtain energy inequalities for a $1$-dimensional signal in $C_n(M)$: 
 Theorem \ref{energyth}.  This special case 
 models a set of $n$ individuals in the Earth atmosphere. In section \ref{sec:graphemb}, considering 
 $n$ points in an arbitrary Riemannian manifold $M$ (not necessarily an open subset of $\R^3$), we approach this issue from the perspective of graph embeddings.  We characterize those via a function that we call the relative ratio function. Informally, its value quantifies how much 
the standard metric on a graph of $n$ vertices differs from the metric induced on the graph when it is identified with a specific configuration of $n$ points in a Riemannian manifold (but when the distance is measured only between the pairs of points that are connected by an edge). The objective is to find a configuration, or a graph 
quasi-embedding, that minimizes this function. We prove Propositions \ref{propfunc1}, \ref{propfunc2}. 

The broader perspective in this paper and in \cite{barron:gs1} is to give an abstract, coordinate-free definition of a signal and related concepts such as energy, noise, or Fourier transform. Generally, abstract mathematical approaches help build the analytical theory, view the systems and processes conceptually,  
 and simplify calculations. To borrow an elementary example from linear algebra: instead of writing a hundred linear equations in a hundred variables, and solving these equations one by one, we can write one equation  $Ax=b$, where $A$ is the matrix of coefficients, $b$ is a vector,  $x$ is the vector of variables, and if $A$ has a nonzero determinant (i.e. columns of $A$ are linearly independent), then we conclude that there is a unique solution $x=A^{-1}b$.  We are pursuing a similar style of thinking in geometry. In practical applicatons, 
 signal processing is, in its essence, numerical analysis on large amounts of data. Because signal propagation typically happens in a space that has some geometry (whether it is a physical space or a parameter space for the system) and processes or propagation are tied with an implied notion of time, we could think most abstractly of a signal as a mapping (a function) between two spaces, that have an appropriate geometric structure. But, by definition, a function from a set $A$ to a set $B$ is a subset of the set $A\times B$. This leads us to defining a signal as a submanifold of a (Riemannian) manifold.  In   \cite{barron:gs1}, instead of allowing any submanifold, we stated that a signal is a cobordism between two submanifolds, thus implicitly taking into account time ( a real parameter associated with evolution).  
Noise was a deformation of the metric. What about energy ? There are various ways to assign a positive real number to a submanifold of a Riemannian manifold, to characterize its "size" or "capacity". For example, one could calculate its volume, but this most obvious approach is not sensitive to the geometric properties of submanifolds. The definition in \cite{barron:gs1}, in particular, uses the cobordism structure of these submanifolds. In the present paper, we allow signals to be arbitrary (bounded) submanifolds or Riemannian manifolds. We still 
expect some notion of process to be intrinsically present (Definition \ref{def:signal}).  
We no longer restrict the considerations to cobordisms, and the time does not have to be "$1$-dimensional". The definitions of noise and energy   from   \cite{barron:gs1} are modified  to this setting.  

Work in the paper is subdivided in the following themes: definition of signals and their energy, along with examples, the space of Gaussians with the product metric, configurations of points, especially in $S^2\times [0,1]$ (or $S^2$, which can be treated the same way). Theorem \ref{th:enbound} establishes estimates on the energy of a signal in a manifold. It applies to the specific cases when the signal is in a manifold of configurations of points or in the  manifold of Gaussians, and in these two cases we also obtain other bounds on energy that are particular to these two settings. On one hand, the discussion of these two types of manifolds can be viewed as an application of our approach to signals and energy. On the other hand, the common underlying theme in this paper is seeking the mathematical methods for describing the set of  $n$ individuals on Earth,  capturing the time evolution and possibly other parameters. The most widely used models in social sciences (e.g. books \cite{hr,wf}) rely on graph theory. Bringing geometry of the ambient space into consideration enriches the approach. Most simply, one can do this via embedding the graph into a metric space or into a manifold (Definitions \ref{metricemb}, \ref{qme}, references \cite{gw:85}, \cite{ren:18}). However, it would be somewhat restrictive to insist that a set of $n$ individuals must be modeled as a graph and treated in the realm of graph theory. We use the manifold of configurations of $n$ points in $S^2\times [0,1]$ to represent all possibilities for the locations of $n$ people at a given moment of time, and we treat, say a path in this manifold as a signal. 
This point of view is deterministic. If, instead, we want to describe one person, say,  not by a point in $\R^3$, but by a probability distribution on $\R^3$, then one approach would be to work with $C_n(D(\R^3))$, where $D(\R^3)$ denotes an appropriate space of probability distributions on $\R^3$ and   $C_n(D(\R^3))$ is the space of configurations of $n$ points in $D(\R^3)$. We do not develop this idea in the present paper, but as a first step  in this direction, we do a calculation on the space of Gaussian distributions on $\R$, taking as a signal a path in the space of Gaussians. The energy of this signal is bounded below in terms of the distance between the endpoints of the path. 
At the end of the paper, we make an effort to tie the graph theory perspective with configurations of points. For a fixed weighted graph 
with $m$ edges 
(that is supposed to describe our set of $n$ individuals), and a Riemannian manifold, with each possible configuration of $n$ points in this manifold, we associate a vector in $\R^m$ and a real number (i.e. we define  a function of $m$ real variables that are components of this vector).   
 To clarify: instead of considering graph embeddings or quasi 
embeddings (i.e. instead of anything in the spirit of Definitions \ref{metricemb}, \ref{qme}), in this setup the graph is fixed, the Riemannian manifold is fixed, and we allow all possible maps from the set of vertices into the manifold but we keep track of how much each map distorts the original structure on the graph (that's the role of the relative ratio function).

\noindent {\bf Acknowledgements.} We are thankful to Rukmini Dey for related communication. We appreciate the valuable suggestions from the referee.   

\section{Signals} 
\label{sec:signals}

Let $(\tilde{M},\tilde{g})$ be a Riemannian manifold.

\begin{defn}
\label{def:signal}
A {\it signal} is $(M,g,A,B)$, where $M$ is  an (embedded) connected submanifold of $\tilde{M}$, with or without boundary/corners \cite{djoyce} 
(we will assume $M\subset \tilde{M}$ and $M$ is embedded via the inclusion map $x\mapsto x$),    
$g$ is the Riemannian metric on $M$ induced by $\tilde{g}$, and $A,B\subset M$ are nonempty closed manifolds such that $A\cap B=\emptyset$. 
If $\partial M\ne \emptyset$, then we assume   $A,B\subset \partial M$. 
\end{defn}
In this paper, we will moreover assume that the closure of $M$ in $\tilde{M}$ is compact. 

This definition differs from the definition of a signal in \cite{barron:gs1}. A detailed discussion follows below.  

Let $(M,g,A,B)$ be a signal. We will write $dV_g$ for the volume form and $\rho_g$ for the distance, in $M$. 
\subsection{Energy}
\begin{defn} 
\label{def:energy}
Let $(M,g,A,B)$ be a signal. We define its {\it $1$-energy} and  {\it $2$-energy}:  
$$
E_1=E_1(M,g,A,B)=\int_M\rho_g(x,A)dV_g(x)
$$
$$
E_2=E_2(M,g,A,B)=\int_M(\rho_g(x,A))^2dV_g(x)
$$

\end{defn}
\begin{example}
\label{exm:lineseg}
Let $\tilde{M}=\R^2$ with the standard Riemannian metric. Let $M$ be the rectangle 
$$
M=\{ (x,y)\in \R^2 | \ -1\le x\le 1; 0\le y\le 1\} ,
$$
$A$ and $B$ be the horizontal line segments 
$$
A=\{ (x,y)\in \R^2 | \ -1\le x\le 1; y=1\} ,
$$
$$
B=\{ (x,y)\in \R^2 | \ -1\le x\le 1; y=0\} .
$$
We calculate:
$$
E_1=\int_M(1-y)dxdy=2\int_0^1(1-y)dy=1
$$
$$
E_2=\int_M(1-y)^2dxdy=2\int_0^1(1-y)^2dy=\frac{2}{3}
$$
\end{example}
\begin{figure}[hbtp]
\includegraphics[width=3.5in]{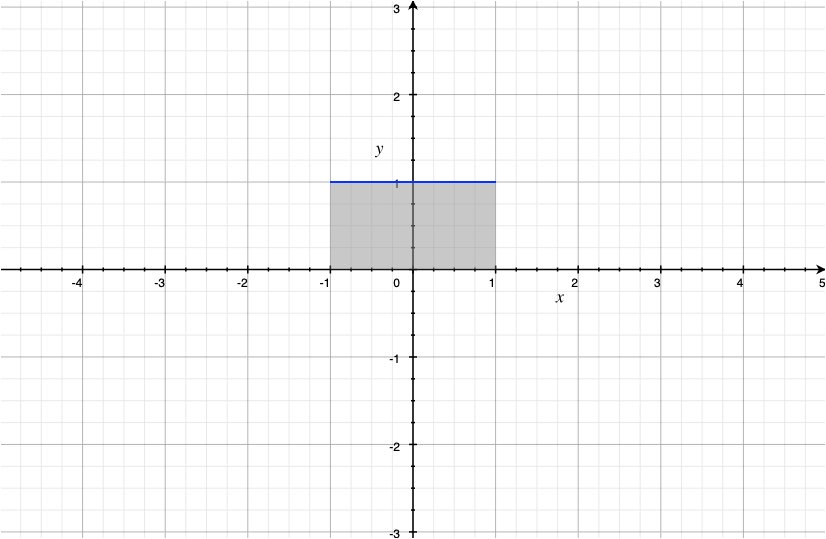}
\caption{$(M,g,A,B)$ in Example \ref{exm:lineseg}.} \label{fig1}
\end{figure} 
The definition of $1$-energy  is similar to the one introduced in  \cite{barron:gs1}. Let's compare different concepts of energy in the following basic example. 
\begin{example} 
\label{ex:fx}
Let $\tilde{M}=\R^2$ with the standard Euclidean metric $\tilde{g}$. Let $M$ be the submanifold (with boundary) which is the graph of a continuously differentiable function $f$ on $[0,3]$, i.e. the set 
$$
M=\{ (x,y)\in\R^2 \ | \ 0\le x\le 3;y=f(x)\}
$$
(see Fig. \ref{fig:fx}).  
\begin{figure}[hbtp]
\includegraphics[width=3.5in]{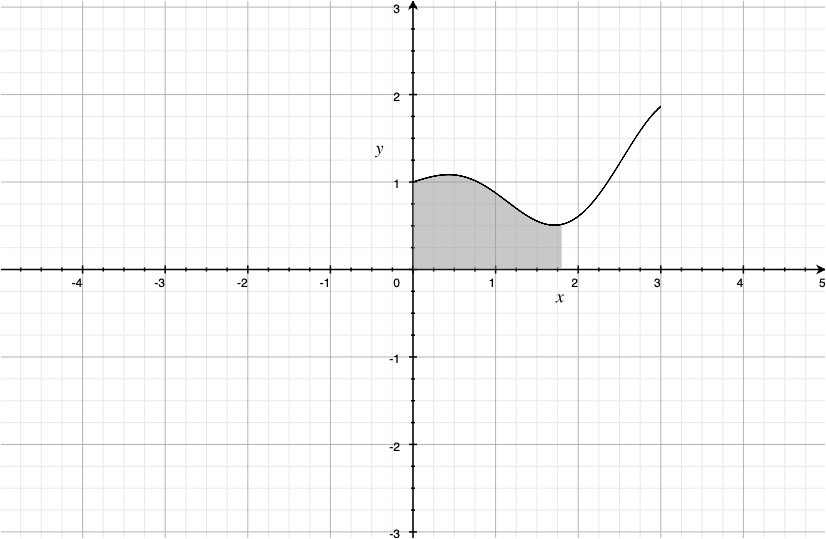}
\caption{$y=f(x)$, $0\le x\le 3$} \label{fig:fx}
\end{figure} 
It is a signal (Definition \ref{def:signal} or \cite{barron:gs1}). Denote the point $(0,f(0))$ by $A$ and the point $(3,f(3))$ by $B$. The energy (Definition \ref{def:energy}) is 
\begin{equation}
\label{eq:en1}
E_1=\int_A^B\rho_g(A,\tau)ds_g(\tau)
\end{equation}
where $\rho_g(A,\tau)$ is the arc length between $A$ and the point $\tau$ on the curve $M$, and 
$$
E_2=\int_A^B\rho_g(A,\tau)^2ds_g(\tau).
$$
Here $g$ is the Riemannian metric on $M$ induced by $\tilde{g}$, $ds_g=dV_g$ is the line element (the volume form for $g$) on $M$, and 
$$
\rho_g(A,\tau)=\int_A^{\tau}ds_g(\lambda).
$$
The Riemannian definition of energy (\cite{jjost} Ch. 8) yields
\begin{equation}
\label{eq:riem}
E_{Riem}=E_{Riem}(f)=\frac{1}{2}\int_0^3(1+f'(x)^2)dx
\end{equation}
and the standard signal processing definition of energy is 
\begin{equation}
\label{eq:sigproc}
E_{sp}=E_{sp}(f)=\int_0^3f(x)^2dx. 
\end{equation}
The signal $M$ can be transformed into another signal, a curve  in $\R^2$, the graph of a function over $0\le x\le 3$, defined by 
$$
F(x)=\int_0^xf(u)du
$$
(the value of $F(x)$ equals the area under the graph of $f$ from $0$ to $x$; the function $F$ is an antiderivative of $f$ on $[0,3]$: $F'(x)=f(x)$). 
The Riemannian energy (\ref{eq:riem}) of the transformed signal is 
$$
E_{Riem}(F)=\frac{1}{2}\int_0^3(1+F'(x)^2)dx=\frac{1}{2}\int_0^3dx+\frac{1}{2}\int_0^3f(x)^2dx=\frac{3}{2}+\frac{1}{2}E_{sp}(f).
$$ 
To relate the definitions (\ref{eq:en1}) and (\ref{eq:sigproc}), we transform the signal $M$ into another signal, a curve  in $\R^2$, the graph of a function over $0\le x\le 3$, defined by 
$$
L(x)=(\int\limits_0^x\sqrt{1+f'(t)^2}dt\sqrt{1+f'(x)^2})^{1/2}.
$$
The energy (\ref{eq:sigproc}) 
of this transformed signal equals the energy (\ref{eq:en1}) of $M$: 
$$
E_{sp}(L)=\int_0^3 L(x)^2dx=E_1(f).
$$
\end{example}
\subsection{More about energy} 
\begin{remark}
As before, let $(\tilde{M},\tilde{g})$ be a Riemannian manifold. An embedded submanifold is a pair $(Q,q)$, where $Q$ is a manifold and $q:Q\to \tilde{M}$ is an embedding. In our setting  
(Definition \ref{def:signal}) $M=q(Q)$, or, instead, we can assume $Q\subset \tilde{M}$ and $q$ is the inclusion map. 
  
  We defined a signal to be a submanifold of $\tilde{M}$ (with appropriate assumptions), rather than a function on this submanifold. In this setting, a signal is a geometric object which is, generally, the image of a map, or, possibly, the graph of a function.     
The function itself plays no role in defining energy of the signal. If we consider $M\subset \tilde{M}$ as the image of itself under the inlcusion map, 
then there is nothing to discuss.  
If $M=q(Q)$, where $Q$ is an abstract smooth manifold (not a subset of $\tilde{M}$), then having only the metric $\tilde{g}$ on $\tilde{M}$, there is  no  notion of a distance between a point of $Q$ and a point of $M$. However, if we modify this (very particular) situation and start with a submanifold $Q$ of $\tilde{M}$ (a subset of $\tilde{M}$, embedded via the inclusion map), and a map $f:Q\to \tilde{M}$ which is a diffeomorphism onto the image, then we could define the energy of such signal as 
\begin{equation}
\label{energyfunc}
\int_Q\rho_{\tilde{g}}(x,f(x))^2d\mu_g(x)
\end{equation}
where $d\mu_g(x)$ is the volume form in the Riemannian metric $g$ on $Q$ induced by $\tilde{g}$.  
In the Example \ref{ex:fx}, with $\tilde{M}=\R^2$ and $Q=[0,3]$, the value of (\ref{energyfunc}) is $E_{sp}$ given by equation (\ref{eq:sigproc}).   
\end{remark}
\begin{remark}
To follow up on the discussion in \cite{barron:gs1} about the signals that occur specifically in speech processing. A word is an ordered finite sequence of symbols (letters). A letter is a submanifold of $\R^2$ (either an image of an appropriate symbol, or the submanifold defined by the soundwave for this letter). Either way, 
for a word of $l$ letters ($l\in\N$) $\Lambda_1$,...,$\Lambda_l$, we get a submanifold 
$\Lambda_1\times ...\times \Lambda_l$
of $\underbrace{\R^2\times ...\times \R^2}_{l}$, with total $1$-energy $E_1(\Lambda_1)+...+E_1(\Lambda_l)$. Similarly for $E_2$. 
\end{remark}
\subsection{Noise, Fourier transform}  
 
For a signal of the form $y=f(t)$, where $f$ is an integrable function on $\R$, the Fourier transform ( a map from $L^1(\R)$ to 
another function space on $\R$) helps identify which frequencies are present 
and the noise is usually defined as the Fourier transform of the autocorrelation function 
$$
a_f(\tau)=\int_{-\infty}^{\infty} f(t+\tau)f(t)dt. 
$$ 
In our definitions, $M$ is bounded in $\tilde{M}$. In \cite{barron:gs1}, we attempted to find "bounded/compact analogues" that would extend, very loosely, the concepts of Fourier transform and noise: as a map $M\to M$ and a local deformation of the metric, respectively. 
\begin{remark}
Suppose $f$ is such that its Fourier transform $F(f)$ has compact support. Let $\beta:\R\to\R$ be a bump function. Let $\varepsilon >0$. Consider 
the Fourier transform of 
$$
a_{f+\varepsilon\beta}(\tau)= \int_{-\infty}^{\infty} (f(t+\tau)+\varepsilon \beta (t+\tau))(f(t)+\varepsilon \beta (t))dt
$$
(i.e. the noise). We observe that as $\varepsilon\to 0$,  the linear approximation of $F(a_{f+\varepsilon\beta})$ has compact support. That, 
in a sense, goes along with our thinking in defining noise as a localized distortion of the metric. 
\end{remark}
\begin{remark}
A more precise "geometric" definition of Fourier transform would require choosing a way to decompose $M$ into "component subsets" and then  defining the FT in such way that it would detect the presence of specifc geometric components (this may be made possible by incorporating Riemannian foliations into the definition).   
\end{remark}
In the present paper, we are not
entering into further discussion of Fourier transform or noise.

  \subsection{Relativity considerations and time}
 
 \subsubsection{} The manifold $\tilde{M}$ is an embedded submanifold of $\tilde{M}\times\R$ via the map 
 $$
 \iota_0:\tilde{M}\to \tilde{M}\times\R
 $$
 $$
 x\mapsto (x,0)
 $$
 Equip $\tilde{M}\times\R$ with the Lorentzian metric $\tilde{g}-\alpha_0dt^2$, where $\alpha_0>0$ is a constant. 
 Standard causal restrictions on the trajectories in $\tilde{M}\times\R$ imply that the physically meaningful paths in $\tilde{M}$ 
 are those with the tangent vectors whose norm does not exceed an upper bound determined by $\alpha_0$.   
 
 In our definition of a signal, "parametrization" or "time" is not explicit, and therefore there is no pressing need to extend the setting to 
 pseudo-Riemannian manifolds. 
 \begin{example}
 For example, the upper arc $M$ of the unit circle in $\R^2$ (with the boundary disjoint union of $p=(1,0)$ and $q=(-1,0)$) is a signal with respect to the Definition \ref{def:signal}. However, a parametrization is not given. The corresponding trajectory in the $3$-dimensional spacetime  
$\R^2\times \R$ is a space curve. Parametrizing the motion (propagation) as $x=\cos t$, $y=\sin t$, $0\le t\le \pi$, we obtain a part of the spiral 
 $$
 x=\cos t;  \ y=\sin t; \  z=t; \ 0\le t\le \pi
 $$
 that is the spacetime trajectory corresponding to $M$. Another parametrization of $M$, e.g. $x=\cos (\pi t^2)$, $y=\sin(\pi t^2)$, $0\le t\le 1$, 
yields a different spacetime trajectory in $\R^2\times\R$ and a different "time" on $M$.    
 \end{example} 
 
\subsubsection{}  Most often, time is assumed to "one-dimensional", meaning that among the coordinates or variables parametrizing the system 
(local or global), there is exactly one that parametrizes the time evolution. In the theory of integrable systems, 
it is common to use multi-time $(t_1,...,t_m)\in \R^m$ for $m\in\N$. 
 In \cite{barron:gs1},  the setup implied intrinsic dynamics or evolution and for this reason the definition of a signal in   \cite{barron:gs1} involved a cobordism. In this paper, we use the definition of signal that allows dynamics with $m$-dimensional time, where $m$ is a positive integer, i.e. 
  time can be parametrized by points of an $m$-dimensional manifold.  
 
 To describe an example of a physical system that is based on such mathematical formalism, we recall that the gravitational time dilation is measured and confirmed by experiments with atomic clocks. Consider a spacecraft departing from Earth in the direction orthogonal to the surface and later returning to the surface. The time $T$ elapsed on clock in the cabin of the spacecraft at the time of the return is the function of the time elapsed on the Earth surface $t$ and of the variables that determine the spacecraft flight plan. For simplicity, we will write 
 $$
 T=T(t,F).
 $$
 where we assume the trajectory of the craft to be determined by one real variable $F$.  
  In general $T\ne t$. It is possible to write an explicit example, with a choice of a specific simple expression for $F$, that yields an explicit formula for $T$.   
  Changing the return time $t$ and changing  $F$ (smoothly), we get that the proper time $T$ is now a function of two variables. 
  The value of $T$, at each $(t,F)$, is one real number. The set 
  $$
  \{ (t,T) \ | \ t>0;T=T(t,F)\} 
  $$
  where $F$ takes values in the appropriate interval in $\R$, consists of pairs of values of two different time parameters, and can be interpreted as "two-dimensional time".

\
 
 This motivates the following definitions.        
    
Let $\tilde{M}$ be a smooth manifold. Let $p,q\in \tilde{M}$ be such that $p\ne q$. A {\it path from $p$ to $q$} is a  $C^1$  map 
$\gamma: [0,1]\to \tilde{M}$ such that $\gamma(0)=p$ and $\gamma(1)=q$. We assume that $\gamma([0,1])$ is an embedded one-dimensional submanifold with boundary.   
We will also call this a {\it one-dimensional path}.

\begin{figure}[hbtp]
\includegraphics[width=3.5in]{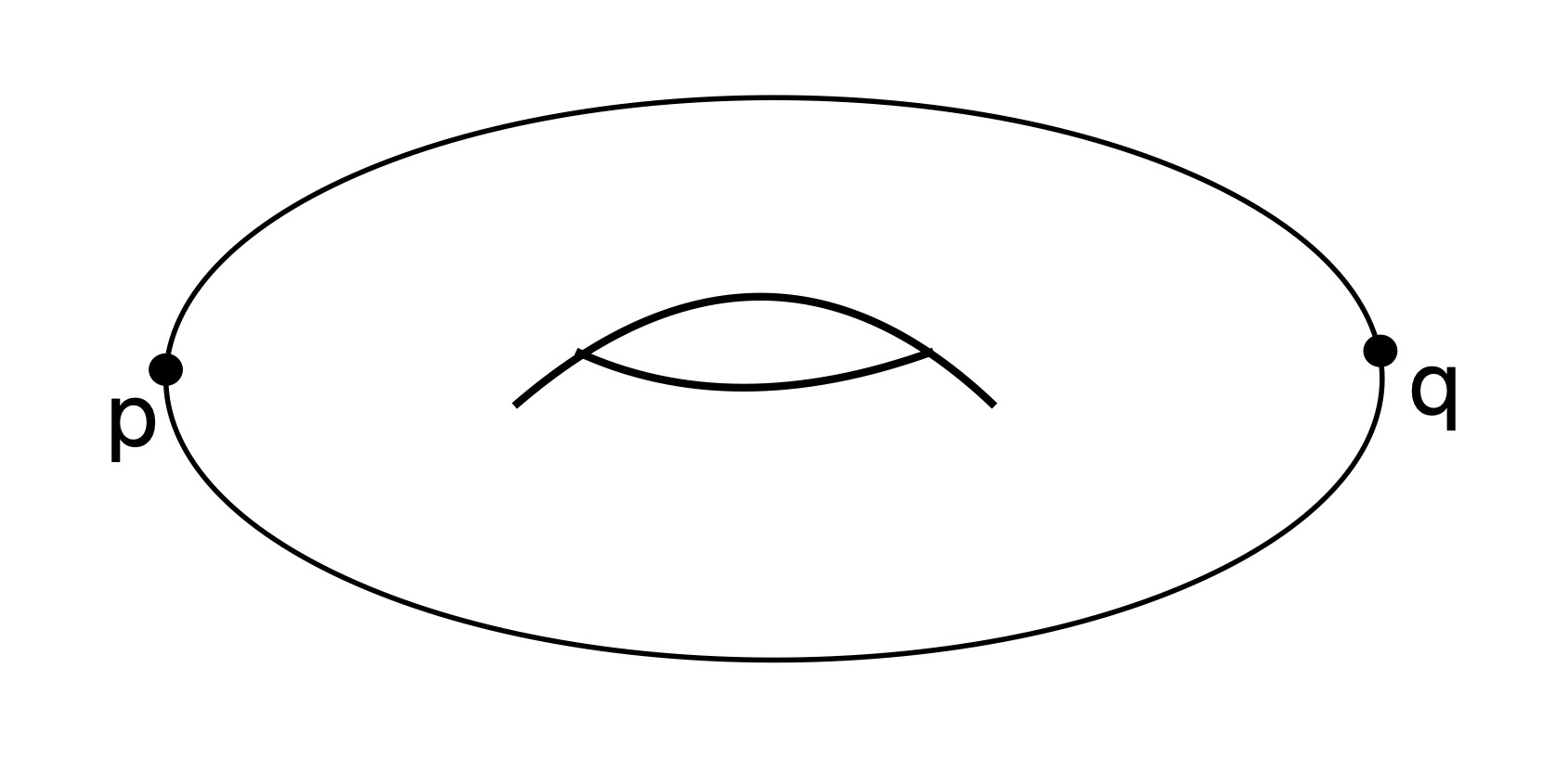}
\caption{A $2$-dimensional path from $p$ to $q$.} \label{fig:torus}
\end{figure}

\begin{defn}
\label{def:kpath}
Let $\tilde{M}$ be a smooth manifold. Let $p,q\in \tilde{M}$ be such that $p\ne q$. Let $k\ge 2$ be an integer. A {\it $k$-dimensional path from $p$ to $q$} is a closed $k$-dimensional manifold $\Lambda$, together with an ordered pair $(a,b)$ of distinct points $a,b\in\Lambda$ and an embedding 
$$
\gamma:\Lambda \to \tilde{M}
$$
such that $\gamma (a)=p$ and $\gamma (b) =q$.    
\end{defn}
\begin{example} 
Let $\tilde{M}=\R^3$. Let $\Lambda$ be the unit sphere in $\R^3$ (embedded via the inclusion map $\gamma:x\mapsto x$)  
$$
\Lambda=\{ x=(x_1,x_2,x_3)\in \R^3 \ | \ x_1^2+x_2^2+x_3^2=1\} .
$$ 
It is a two-dimensional path from $p=(-1,0,0)$ to $q=(1,0,0)$. 
\end{example} 
Another example of a two-dimensional path in $\R^3$ is shown in Figure \ref{fig:torus}. 

We observe that according to the Definition \ref{def:energy}, the $1$-energy and $2$-energy of a $k$-dimensional path 
from $p$ to $q$ in a Riemannian manifold $(\tilde{M},\tilde{g})$ are 
$$
E_1=\int_M\rho_g(x,p)dV_g(x)
$$
and
$$
E_2=\int_M\rho_g(x,p)^2dV_g(x),
$$  
where $M=\gamma([0,1])$ when $k=1$, $M=\gamma(\Lambda)$ when $k\ge 2$, $g$ is the Riemannian metric on $M$ induced by $\tilde{g}$. 
\begin{remark}
We have already explained above that this is not the standard Riemannian geometry definition of the energy of a path, but since this number quantifies the strength, or capacity of the signal, we call it energy for the purposes of this discussion.  
\end{remark}
\begin{theorem}
\label{th:enbound}
Let $(\tilde{M},\tilde{g})$ be a Riemannian manifold. 

\noindent $(i)$ \  Let $\gamma:[0,1]\to \tilde{M}$ be a one dimensional path from $p=\gamma(0)$ to $q=\gamma(1)$. 
For the signal 
$(M=\gamma([0,1]),g,p,q)$ 
$$
E_1\le \rho_g(p,q)^2; \  E_2\le \rho_g(p,q)^3.
$$     
\noindent $(ii)$ \ Let $k\ge 2$. 
Let $k,a,b,\Lambda,\gamma,p,q$ define a $k$-dimensional path from $p$ to $q$ in $\tilde{M}$, as in Definition \ref{def:kpath}. For the signal 
$(M=\gamma(\Lambda),g,p,q)$ 
$$
E_1\le diam_g(M) vol_g(M)
$$     
$$
E_2\le diam_g(M)^2 vol_g(M).
$$
\end{theorem}
\noindent {\bf Proof.} 
To prove $(i)$: for each point $x\in M$ 
$$
\rho_g(x,p)\le \rho_g (p,q) ,
$$
therefore
$$
E_1=\int_M\rho_g(x,p)dV_g(x)\le \int_M\rho_g(p,q)dV_g(x)= \rho_g(p,q)^2
$$
and
$$
E_2=\int_M\rho_g(x,p)^2dV_g(x)\le \int_M\rho_g(p,q)^2dV_g(x)= \rho_g(p,q)^3.
$$
For  $(ii)$, we note that  
for each point $x\in M$ 
$$
\rho_g(x,p)\le diam_g(M) ,
$$
it follows that
$$
E_1=\int_M\rho_g(x,p)dV_g(x)\le \int_M diam_g(M)dV_g(x)= diam_g(M)vol_g(M)
$$
and
$$
E_2=\int_M\rho_g(x,p)^2dV_g(x)\le \int_M diam_g(M)^2dV_g(x)= diam_g(M)^2vol_g(M).
$$
$\Box$

\section{Parameter spaces of Gaussians}
\label{sec:gau}
Let $n\in\N$. Let $P_n$ be the manifold of positive definite $n\times n$ real matrices, with the standard Riemannian 
metric induced from the inner product on the space of $n\times n$ real matrices defined by 
$$
(A,B)=\tr(A^T B)
$$
(see Chapter 6 \cite{posdef}). Let $\Omega\subset \R^n$ be a domain (i.e. an open connected set). Let 
$$
\mathcal{I}=\Omega \times P_n=\{ (\mu,\Sigma) \ | \  \mu\in\Omega;  \Sigma\in P_n\} . 
$$
The manifold $\mathcal{I}$ parametrizes Gaussian distributions on $\R^n$:
$$
f(x)=\frac{1}{\sqrt{(2\pi)^n\det \Sigma}}e^{ -\frac{1}{2}(x-\mu)^T\Sigma^{-1}(x-\mu)}.
$$
In information geometry, $\mathcal{I}$ equipped with the Fisher metric, is a statistical manifold. We will instead assume 
$\Omega\times P_n$ is endowed with the product metric. On $\R^n\times P_n$, the product metric is the Euclidean metric, and 
with the standard choice of coordinates 
$(x_1,...,x_N)$, where $N=n+\frac{n(n+1)}{2}$, the distance between points $a$ and $b$ in $\R^n\times P_n$ equals 
$$
||a-b||_2=\sqrt{(a_1-b_1)^2+...+(a_N-b_N)^2}.
$$
Also, recall that for a real number $p\ge 1$, the distance between $a$ and $b$ in the $L^p$ norm is 
$$
||a-b||_p=\Bigl (|a_1-b_1|^p+...+|a_N-b_N|^p\Bigr ) ^{1/p}.
$$
\begin{example} 
To illustrate the difference between the Fisher metric and the product metric: let $n=1$, $\Omega=\R$.  The manifold $\mathcal{I}=\Omega \times P_1$ parametrizes Gaussian distributions on $\R$:
$$
f(x)=\frac{1}{\sqrt{2\pi\sigma^2}}e^{ -\frac{(x-\mu)^2}{2\sigma^2}}.
$$
Here $\mu\in \Omega=\R$ and $\sigma\in P_1\simeq (0,\infty )$. The standard product metric on $\Omega\times P_1$ is 
$$
ds^2_{st}=d\mu^2+d\sigma^2.
$$
Now, let's calculate the components of the metric tensor $(g_{ij})$ for the Fisher information metric $ds^2_{Fisher}$: 
$$
g_{11}=-\int_{\R}   \frac{1}{\sqrt{2\pi\sigma^2}}e^{ -\frac{(x-\mu)^2}{2\sigma^2}}\frac{\partial^2}{\partial \mu^2} 
\ln \Bigl ( \frac{1}{\sqrt{2\pi\sigma^2}}e^{ -\frac{(x-\mu)^2}{2\sigma^2}}\Bigr ) dx=\frac{1}{\sigma^2}
$$
$$
g_{12}=g_{21}=-\int_{\R}   \frac{1}{\sqrt{2\pi\sigma^2}}e^{ -\frac{(x-\mu)^2}{2\sigma^2}}\frac{\partial^2}{\partial \sigma \ \partial \mu} 
\ln \Bigl ( \frac{1}{\sqrt{2\pi\sigma^2}}e^{ -\frac{(x-\mu)^2}{2\sigma^2}}\Bigr ) dx=0
$$
$$
g_{22}=-\int_{\R}   \frac{1}{\sqrt{2\pi\sigma^2}}e^{ -\frac{(x-\mu)^2}{2\sigma^2}}\frac{\partial^2}{\partial \sigma^2} 
\ln \Bigl ( \frac{1}{\sqrt{2\pi\sigma^2}}e^{ -\frac{(x-\mu)^2}{2\sigma^2}}\Bigr ) dx=\frac{2\sqrt{2}}{\sigma^2\sqrt{\pi}} .
$$
Therefore
$$
ds^2_{Fisher}=\frac{1}{\sigma^2}d\mu^2+\frac{2\sqrt{2}}{\sigma^2\sqrt{\pi}} d\sigma^2.
$$
\end{example}
The manifold $\mathcal{I}$ with the product metric is 
$(\tilde{M},\tilde{g})$ of section \ref{sec:signals}. 
\begin{theorem}
\label{th:e2}
Let $n\in\N$. Let $\Omega\subset \R^n$ be a domain. Let $p,q\in \mathcal{I}=\Omega\times P_n$ be such that $p\ne q$. 
Let $\gamma:[0,1]\to \mathcal{I}$ be a one-dimensional path in $\mathcal{I}$ such that $\gamma(0)=p$, $\gamma(1)=q$, 
the convex hull of $\gamma([0,1])$ in $\R^n\times P_n$ is contained in $\mathcal{I}$, and each component function $x_i$ of $\gamma$ is monotone.  For the signal 
$(M=\gamma([0,1]),g,p,q)$
$$
E_2\ge \frac{1}{3}||q-p||_3^3.
$$
\end{theorem}
{\bf Proof.}  We have:
$$
E_2=\int_0^1\rho_g(p,\gamma(t))^2|\gamma'(t)|dt
$$
where $|\gamma'(t)|=||\gamma'(t)||_2$. 
Since for each $t$ 
$$
\rho_g(p,\gamma(t))\ge \rho_{\tilde{g}}(p,\gamma(t))
$$
and for each $i\in \{ 1,...,N\}$
$$
|\gamma'(t)|\ge |x_i'(t)|,
$$
we get: 
$$
E_2\ge \sum_{i=1}^N\int_0^1 (x_i(t)-p_i)^2|x'_i(t)|dt=\sum_{i=1}^N|\int_{p_i}^{q_i} (x_i-p_i)^2dx_i|=
$$
$$
\sum_{i=1}^N\frac{1}{3} |q_i-p_i|^3=\frac{1}{3}||q-p||_3^3.
$$
$\Box$

\section{Configurations of points}
\label{sec:completeg}

\begin{defn}
Let $M$ be a smooth manifold. Let $n\ge 2$ be an integer. The set 
$$
C_n(M)=\{ (x_1;...;x_n)\in \underbrace{M\times ...\times M}_{n} \ | \ x_1,...,x_n\in M; \ x_i\ne x_j \ {\mathrm{for \ all}} \ i\ne j\} \ 
$$
(the set of ordered $n$-element subsets of $M$) is  the {\it set of (ordered) configurations of $n$ points in $M$}.
\end{defn}
There is a vast amount of literature about these spaces, including the fundamental paper \cite{atiyah} about the configurations spaces of $n$ distinct ordered points in $\R^3$. 

For each $j\in \{ 1,...,n\}$, denote by $p_j: \underbrace{M\times ...\times M}_{n}\to M$ the projection onto the $j$-th component: 
$$
p_j:(x_1;...;x_n)\mapsto x_j .
$$

Let $a,b\in\R$ be such that $0<a<b$. Let 
$$
S=\{ v=(x,y,z)\in\R^3 \ | \ a<x^2+y^2+z^2<b\} .
$$

We consider $S$ as a Riemannian manifold, with the Riemannian metric induced by the standard Euclidean metric on $\R^3$. It is also a metric space with the distance function defined by this Riemannian metric. 
Write 
$$
C_n(S)=\{ (v^{(1)};...;v^{(n)})\in \underbrace{S\times ...\times S}_{n} \ | \ v^{(1)},...,v^{(n)}\in S; \ v^{(i)}\ne v^{(j)} \ {\mathrm{for \ all}} \ i\ne j\} \ 
$$
(the set of ordered $n$-element subsets of $S$, the space of configurations of $n$ points in $S$).

\begin{lemma}Let $n\ge 2$ be an integer. Let $X$ be a metric space. Let $i,j$ be integers such that $1\le i<j\le n$. The set  
$$
D_{ij}=\{x=(x_1;...;x_n)\in \underbrace{X\times ...\times X}_{n}| \ x_i=x_j\} 
$$
is closed in $\underbrace{X\times ...\times X}_{n}$.
\end{lemma}
\noindent {\bf Proof.} Let $(c^{(m)})$ be a sequence in $D_{ij}$ that converges to $c\in X$. Since the projections onto the $i$-th factor 
and onto the $j$-th factor 
$p_i,p_j: \underbrace{X\times ...\times X}_{n}\to X$ are continuous maps, 
and for each $m$ 
$$
c^{(m)}_i=c^{(m)}_j, 
$$
it follows that 
$$
p_i(\lim_{m\to\infty}c^{(m)})=\lim_{m\to\infty}p_i(c^{(m)})= \lim_{m\to\infty}p_j(c^{(m)}=p_j(\lim_{m\to\infty}c^{(m)})
$$
i.e. $c_i=c_j$, and hence $c\in D_{ij}$. The statement follows. $\Box$

\begin{lemma}
Let $M$ be a connected Riemannian manifold. Let $n\ge 2$ be an integer. The set 
$C_n(M)$ is a Riemannian manifold, with the  Riemannian metric unduced by 
the Riemannian metric on  $\underbrace{M\times ...\times M}_{n}$. 
\end{lemma} 
\noindent {\bf Proof.} Since an open subset of a manifold is a manifold, it is sufficient to show that $C_n(M)$ is open in 
$\underbrace{M\times ...\times M}_{n}$. 
By the previous Lemma, for each $1\le i<j\le n$ the set 
$$
D_{ij}=\{x=(x_1;...;x_n)\in \underbrace{M\times ...\times M}_{n}| \ x_i=x_j\} 
$$
is closed in $\underbrace{M\times ...\times M}_{n}$. Since 
$$
C_n(M)=\underbrace{M\times ...\times M}_{n}-\bigcup D_{ij},
$$
it follows that $C_n(M)$ is open. $\Box$

\subsection{Riemannian metric}
The standard Riemannian metric on $\R^3$ induces a Riemannian metric on $S$. The product Riemannian metric on 
$\underbrace{S\times ...\times S}_{n}$ gives a Riemannian metric $g$ on the submanifold $C_n(S)$. 
Using the properties of the standard topology on $\R^3$, we deduce that $C_n(S)$ is path-connected.  

\begin{remark} Another way to describe $g$ is as follows. Let $(e_1,e_2,e_3)$ be a standard basis in $\R^3$. Consider the bijection 
$$
\mu_n:\underbrace{\R^3\times ...\times \R^3}_{n}\to Mat_{3\times n}(\R)
$$
where the columns of the $3\times n$ matrix are the coordinate vectors with respect to the basis $(e_1,e_2,e_3)$. 

    Let $V\in C_n(S)$. Let $\gamma_1$, $\gamma_2$ be two smooth paths in $C_n(S)$ through $V$: 
$$
\gamma_1: [-1,1]\to C_n(S); \ \gamma_1(0)=V
$$
$$
\gamma_2: [-1,1]\to C_n(S); \ \gamma_2(0)=V
$$
The Riemannian metric $g$ is defined by 
$$
g_V(\gamma_1'(0),\gamma_2'(0))= \tr \Bigl (  \mu_n(\gamma_1'(0))^T\mu_n(\gamma_2'(0))\Bigr) =
\tr \Bigl (  \mu_n(\gamma_2'(0))^T\mu_n(\gamma_1'(0))\Bigr) .
$$ 
\end{remark}
\subsection{Energy}
Let $n\in\N$. Let 
$A=(a^{(1)};...;a^{(n)})\in C_n(S)$ and $B=(b^{(1)};...;b^{(n)})\in C_n(S)$ be such that $A\ne B$. Let 
$$
\gamma:[0,1]\to C_n(S)
$$
be a smooth path such that $\gamma(0)=A$ and $\gamma(1)=B$. 
Write
$$
\gamma=(\gamma^{(1)},...,\gamma^{(n)})
$$
where $\gamma^{(j)}=p_j\circ \gamma$. 
For each $j$, the function $\gamma^{(j)}$ is a function $[0,1]\to\R^3$. Write $\gamma^{(j)}=(\gamma^{(j)}_1,\gamma^{(j)}_2,\gamma^{(j)}_3)^T$. 
Assume, for each $j$,  the functions $\gamma^{(j)}_k:[0,1]\to\R$, $k=1,2,3$, are monotonic. 
As  in section \ref{sec:signals}, we consider 
$M=\gamma([0,1])\subset C_n(S)$ to be a signal and define its {\it 1-energy} as  
\begin{equation}
\label{eq:e1}
E_1=\int_0^1\rho(A,\gamma(t))|\gamma'(t)|dt
\end{equation}
where $\rho$ is distance with respect to the Riemannian metric induced on $M$ by $g$, and {\it  2-energy} by 
\begin{equation}
\label{eq:e2}
E_2=\int_0^1(\rho(A,\gamma(t))^2|\gamma'(t)|dt. 
\end{equation}
We also write, for each $j\in \{1,...,n\}$  
\begin{equation}
\label{eq:e1j}
E_1^{(j)}=\int_0^1\rho_j(a^{(j)},\gamma^{(j)}(t)|(\gamma^{(j)}(t))'|dt
\end{equation}
where $\rho_j$ is the distance on the curve $\gamma^{(j)}:[0,1]\to S$ induced by the standard Riemannian metric on $\R^3$ and 
\begin{equation}
\label{eq:e2j}
E_2^{(j)}=\int_0^1(\rho_j(a^{(j)},\gamma^{(j)}(t))^2|(\gamma^{(j)}(t))'|dt.
\end{equation}
\begin{theorem} 
\label{energyth}
\noindent $(i)$ \ $E_1\le \rho(A,B)^2$; \  $E_2\le \rho(A,B)^3$

\noindent $(ii)$ \ For each $j\in \{ 1,...,n\}$ 
$$
E_1\ge E_1^{(j)}
$$
and 
$$
E_2\ge E_2^{(j)}.
$$
\noindent $(iii)$ \ 
Suppose $\gamma:[0,1]\to C_n(S)$ be a one-dimensional path in $C_n(S)$ such that $\gamma(0)=A$, $\gamma(1)=B$, 
the convex hull of $\gamma([0,1])$ in $\underbrace{\R^3\times ...\times \R^3}_{n}$ is contained in $C_n(S)$, and each 
component function 
$\gamma^{(j)}_k$ ($j\in \{1,...,n\}$, $k\in \{ 1,2,3\}$) 
of $\gamma$ is monotone.  For the signal 
$(M=\gamma([0,1]),g,A,B)$
$$
E_2\ge \frac{1}{3}||B-A||_3^3.
$$
\end{theorem}
{\bf Proof.} Proof of $(i)$. 
The function $\rho(A,\gamma(t))$ regarded as an $\R$-valued function on $[0,1]$ is increasing: 
 $\rho(A,\gamma(t))\le \rho (A,B)$ for each $t$. Hence 
$$
E_1=\int_0^1\rho(A,\gamma(t))|\gamma'(t)|dt\le \rho(A,B)\int_0^1|\gamma'(t)|dt=\rho(A,B)^2.
$$
The proof of the second inequality is similar.  

To prove $(ii)$, we use that for each $t$ and each $j\in \{ 1,...,n\}$ 
$$
|\gamma(t)'|\ge |\gamma^{(j)}(t)'|
$$
and
$$
\rho(A,\gamma(t)) \ge \rho_{j}(a_j,\gamma^{(j)}(t)).
$$
The inequalities now follow from (\ref{eq:e1}),  (\ref{eq:e2}), (\ref{eq:e1j}), (\ref{eq:e2j}). 

Proof of $(iii)$ is similar to the proof of Theorem \ref{th:e2}. 
 $\Box$

\section{Graph embedddings and relative ratio variance function} 
\label{sec:graphemb}

 Let $\Gamma=(V,E)$ be a simple graph, where 
$V=\{ v_1,...,v_n\}$  is the set of vertices
and 
$$
E\subset \{ \{ v_i,v_j\}\ | \ 1\le i<j\le n\}
$$ 
is the set of edges. 

Assume $\Gamma$ is connected. Define the metric $d_{\Gamma}:V\times V \to \Z$ by setting $d_{\Gamma}(x,y)$ for $x,y\in V$ to be the number of edges in a shortest path  between 
$x$ and $y$. 

\begin{defn} \cite{gw:85, ren:18} \label{metricemb}
Let $M$ be a Riemannian manifold. We say a map $f:V\to M$ is a {\it isometric embedding of $\Gamma$ into $M$} 
if for all $x,y\in V$ 
$$
d_M(f(x),f(y))=d_{\Gamma}(x,y),
$$
where $d_M$ is the distance in $M$ induced by the Riemannian metric. 
\end{defn}
\begin{example}
Consider an undirected graph $\Gamma$ with two vertices and one edge: $V=\{ v_1,v_2\}$, $E$ is a one element set. Let $M=\R^2$ with the standard Euclidean metric. The space of all isometric embeddings of $\Gamma$ into $M$ in the sense of Definition \ref{metricemb} is homeomorphic to $\R^2\times S^1$. 
\end{example}

\begin{defn} \label{qme}
Let $M$ be a Riemannian manifold. We say a map $f:V\to M$ is a {\it quasi-isometric embedding of $\Gamma$ into $M$} 
if for every edge $\{ v_i,v_j\}\in E$ 
$$
d_M(f(v_i),f(v_j))=1,
$$
where $d_M$ is the distance in $M$ induced by the Riemannian metric. 
\end{defn}
This is not the same as the definition \ref{metricemb}. 

Now, let $n\ge 2$ be an integer. Let $\Gamma=(V,E)$ be a graph as above, with $n$ vertices, endowed with a positive weight function 
$$
W:E\to \R.
$$
Let $|E(\Gamma)|=m$. We choose and fix a bijection $\iota_E:\{ 1,...,m\}\to E$. 

Let $M$ be a connected Riemannian manifold of dimension $\dim M\ge 2$.  
Let $C_n(M)$ be the manifold of configuration of $n$ (distinct, ordered)  points in $M$. 
We are now going to define a function whose purpose is to measure how well a given configuration represents a fixed weighted graph. 
Let $X\in C_n(M)$. For each $k\in \{ 1,...,m\}$ denote  
$$
r_k=\frac{d_M(p_i(X), p_j(X))}{W(v_i,v_j)} 
$$
where $\{ v_i,v_j\}\in E$ is $\iota_E(k)$.
Define the function 
$$
\tilde{v}:C_n(M)\to\R
$$
by $\tilde{v}=v\circ r$, where 
$$
r:C_n(M)\to\R^m
$$
is defined by $r(X)=(r_1,...,r_m)$ as above, and $v:\R^m\to\R$ is defined by 
$$
v(r_1,...,r_m)=\frac{\sum\limits_{i=1}^{m} (r_i-\frac{r_1+...+r_{m}}{m})^2 } { \frac{(r_1+...+r_m)^2}{m^2} }.
$$
\begin{lemma} 
The function $\tilde{v}$ is continuous. 
\end{lemma} 
{\bf Proof.} Since all $r_k>0$, it is immediately clear that $v$ is continuous. It is straightforward to verify that each component of $r$ is continuous. Therefore $r$ is continuous. It follows that the composition is continuous. $\Box$ 
\begin{proposition}
\label{propfunc1}
The function $\tilde{v}$ attains a minimum value of zero at a configuration $X$ if and only if $r(X)=(r_1,...,r_m)$ is such that 
$r_1=...=r_m$. 
\end{proposition} 
\noindent {\bf Proof}. Suppose  $\tilde{v}(X)=0$. Then for $r(X)=(r_1,...,r_m)$ we have:
$$
\sum\limits_{i=1}^{m} (r_i-\frac{r_1+...+r_{m}}{m})^2=0.
$$
It follows that 
$$
r_1+...+r_m=mr_i
$$
for each $i\in \{ 1,...,m\}$. Therefore all $r_i$ are equal. 

Conversely, if  for $r(X)=(r_1,...,r_m)$, $r_1=...=r_m$, then for each $i\in \{ 1,...,m\}$ 
$$
r_i-\frac{r_1+...+r_{m}}{m}=0
$$
and hence $\tilde{v}(X)=0$. Since $\tilde{v}(X)$ is always nonnegative, the conclusion follows. $\Box$
\begin{proposition}
\label{propfunc2}
Let $\alpha>0$ be a positive constant. Suppose $f:M\to M$ is a  homeomorphism of $M$ such that 
$$
d_M(f(x),f(y))=\alpha d_M(x,y)
$$
for all $x,y\in M$. 
The function $\tilde{v}$ is $f$-invariant in the following sense:
let for each $k\in \{ 1,...,m\}$ 
$$
\hat{r}_k=\frac{d_M((f\circ p_i)(X), (f\circ p_j)(X))}{W(v_i,v_j)} ,
$$
then 
\begin{equation}
\label{eq:invar}
(v\circ \hat{r}) (X)=\tilde{v}(X).
\end{equation}
\end{proposition} 
\noindent {\bf Proof.} 
Verifying (\ref{eq:invar}) is, in its essence, the following simple observation: 
$$
(v\circ \hat{r}) (X)=
\frac{\sum\limits_{i=1}^{m} (\alpha r_i-\alpha\frac{r_1+...+r_{m}}{m})^2 } { \frac{(\alpha r_1+...+\alpha r_m)^2}{m^2} }=
\frac{\sum\limits_{i=1}^{m} (r_i-\frac{r_1+...+r_{m}}{m})^2 } { \frac{(r_1+...+r_m)^2}{m^2} }=
\tilde{v}(X).
$$
$\Box$

\end{document}